\documentclass[
   ,final            
  ]{aipproc}
\layoutstyle{6x9}

\def\apj{ApJ}

\def\aap{A\&A}

\def\aaps{A\&AS}

\def\saxsource{SAX~J0840.7$+$2248}
\def\grbsource{GRB~980429}

\def\arcmin{\hbox{$^\prime$}}
\def\arcsec{\hbox{$^{\prime\prime}$}}
\def\utw{\smash{\rlap{\lower5pt\hbox{$\sim$}}}}
\def\udtw{\smash{\rlap{\lower6pt\hbox{$\approx$}}}}

\begin{document}
\title[GRB~980429]{Swift uncovers that SAX~J0840.7$+$2248 is not an X-ray Binary, but BeppoSAX X-ray Rich GRB~980429}

\author{P.~Romano}{address={INAF--Osservatorio Astronomico di Brera, Via E.\ Bianchi 46, I-23807 Merate (LC), Italy},
	altaddress={Universit\`a{} degli Studi di Milano, Bicocca, Piazza delle Scienze 3, I-20126 Milano, Italy}}
\author{C.\ Guidorzi}{address={INAF--Osservatorio Astronomico di Brera, Via E.\ Bianchi 46, I-23807 Merate (LC), Italy},
	altaddress={Universit\`a{} degli Studi di Milano, Bicocca, Piazza delle Scienze 3, I-20126 Milano, Italy}}
\author{L.~Sidoli}{address={INAF--Istituto di Astrofisica Spaziale e Fisica Cosmica, Via E.\ Bassini 15, I-20133 Milano, Italy}}
\author{E.~Montanari}{address={Dipartimento di Fisica, Universita' di Ferrara, via Saragat 1, I-44100 Ferrara},
		altaddress={I.I.S.\ "Calvi", Finale Emilia, Modena}}
\author{F.~Capitanio}{address={INAF--Istituto di Astrofisica Spaziale e Fisica Cosmica,  Via del Fosso del Cavaliere 100, I-00133 Roma, Italy}}
\author{L.~Amati}{address={INAF--Istituto di Astrofisica Spaziale e Fisica Cosmica, Via Gobetti, 101 I-40129 Bologna, Italy}}
\author{A.~Cucchiara}{address={Dept.\ of Astronomy \& Astrophysics, Pennsylvania State University, 
                   University Park, PA 16802}}
\author{F.~Frontera}{address={Dipartimento di Fisica, Universita' di Ferrara, via Saragat 1, I-44100 Ferrara},
		altaddress={INAF--Istituto di Astrofisica Spaziale e Fisica Cosmica, Via Gobetti, 101 I-40129 Bologna, Italy}}
\author{N.~Masetti}{address={INAF--Istituto di Astrofisica Spaziale e Fisica Cosmica, Via Gobetti, 101 I-40129 Bologna, Italy}}
\author{S.~Mereghetti}{address={INAF--Istituto di Astrofisica Spaziale e Fisica Cosmica, Via E.\ Bassini 15, I-20133 Milano, Italy}}
\author{F.~Rossi}{address={Dipartimento di Fisica, Universita' di Ferrara, via Saragat 1, I-44100 Ferrara}}

%
\begin{abstract}
During our Swift/XRT program to obtain X-ray positions
at arcsecond level for a sample of Galactic
X-ray binaries, we discovered that \saxsource{} 
is not a binary, but rather BeppoSAX/WFC+GRBM X-ray Rich
\grbsource. Here we report on this discovery and on the
properties of this long, X-ray rich gamma-ray burst,
from prompt to (very) late followup.
\end{abstract}
\maketitle

\section{Introduction}

Several catalogued X--ray binaries still have sky positions measured with
uncertainties at arcminute level. Such large error boxes prevent 
a fruitful multiwavelength study; in particular, they make it impossible to 
establish a firm association with optical/IR/radio counterparts. 
Therefore, we have a standing Swift fill-in target 
proposal to observe a sample of such objects drawn from the catalogues of 
\cite{Liu2001:lmxbcat,Liu2006:hmxbcat}. 
Among them is the Low Mass X--ray Binary \saxsource,
which was at first classified as a so-called ``burst-only''  source  
\cite{Cocchi2001,Cornelisse2002}. 
It was  observed only once \cite{Heise:1998IAUC6892} 
through a bright ($\sim 1$ Crab peak intensity in the 2--25 keV energy range),
$\sim 100$\,s long burst with the Wide Field Cameras \cite[WFC,][]{Jager1997:SAXWFC} 
on-board BeppoSAX \cite{Boella1997:SAX}, at the position 
RA(J2000$)=08^{\rm h}$ $40^{\rm m}$ $40^{\rm s}$, 
Dec(J2000$)=+22^{\circ}$ $48\arcmin$ $18\arcsec$ (error radius 3\arcmin{}). 
Here we describe the discovery, based on Swift data and reported 
in \cite{Sidoli:2007ATel1089}, that \saxsource{} was actually a gamma ray burst, 
\grbsource.

Throughout this paper the quoted uncertainties are given at 90\% confidence level
for one interesting parameter unless otherwise stated.
We use $\Gamma$  as the power-law photon index,
$N(E) \propto E^{-\Gamma}$ (ph keV$^{-1}$ cm$^{-2}$ s$^{-1}$).

\section{The Swift data\label{xrtdata}}

\begin{figure}
  \includegraphics[height=.55\textheight,angle=270]{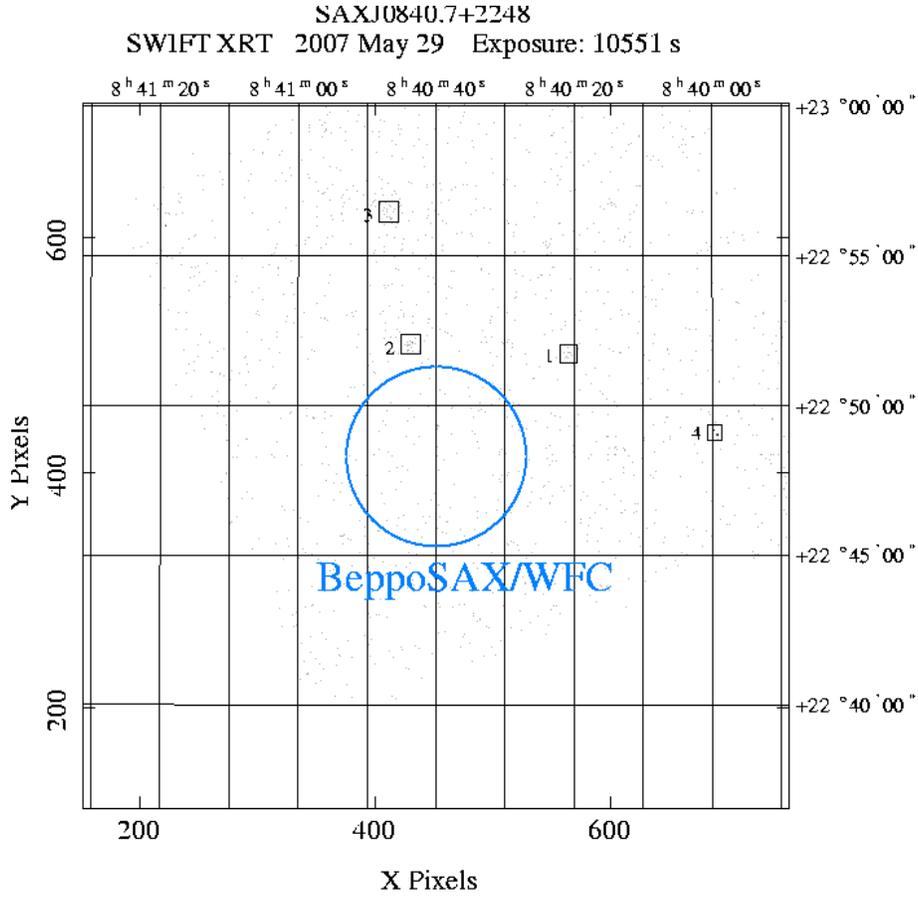}
  \caption{Swift/XRT image of the field of \saxsource, 
	obtained from the total $\sim10.6$\,ks PC mode data. 
	The large circle is the  BeppoSAX/WFC error circle (3$\arcmin$ radius).
	The small squares mark X-ray sources detected at $>3$-$\sigma$ level.
	\label{grb980429:fig_xrt} }
\end{figure}

The Swift \cite{Gehrels04long} data on \saxsource{} were collected as a 
fill in target observation. The source was observed for a total of 10.6\,ks 
on 2007 May 29 00:35:49 to 23:21:57 UT. 
The XRT data were processed with standard procedures ({\tt xrtpipeline} 
v0.10.6), filtering and screening criteria by using {\tt FTOOLS} in the
{\tt Heasoft} package (v.6.1.2). 
The UVOT data were reduced with the standard tools within the same 
{\tt Heasoft} package. 

Figure~\ref{grb980429:fig_xrt} 
shows the field of \saxsource{} as imaged by 
Swift/XRT in photon counting (PC) mode. The data show no X-ray counterpart 
within the BeppoSAX/WFC position error circle (3$\arcmin$ radius) 
centered on RA(J2000$)=08^{\rm h}$ $40^{\rm m}$ $40^{\rm s}$, 
Dec(J2000$)=+22^{\circ}$ $48\arcmin$ $18\arcsec$ \cite{Heise:1998IAUC6892}.   
A 3-$\sigma$ upper limit on a source within the BeppoSAX error box can be   
placed at $4.5\times 10^{-14}$ erg cm$^{-2}$ s$^{-1}$ 
(assuming a spectrum of $\Gamma=2$, no intrinsic absorption,
and a Galactic Hydrogen column of $N_{\rm H} =3.45 \times 10^{20}$ cm$^{-2}$).   
Assuming a distance of 8\,kpc (or 1 kpc), we obtain a 3-$\sigma$ upper 
limit on the luminosity of $3\times10^{32}$ $(5\times 10^{30})$ erg s$^{-1}$.   
These values are quite low for a Galactic X-ray binary hosting a neutron 
star, while black-hole X--ray novae in quiescence
can reach luminosities below 10$^{32}$ erg s$^{-1}$ (e.g.\ \citealt{Kong2002}).
Since the putative source should be an X--ray burster, and
since the UVOT images revealed no new sources in the BeppoSAX/WFC error 
circle, we considered that this event might not be associated with an X-ray binary 
at all.

\section{The BeppoSAX data\label{saxdata}}

\begin{figure}
  \includegraphics[height=.55\textheight,angle=0]{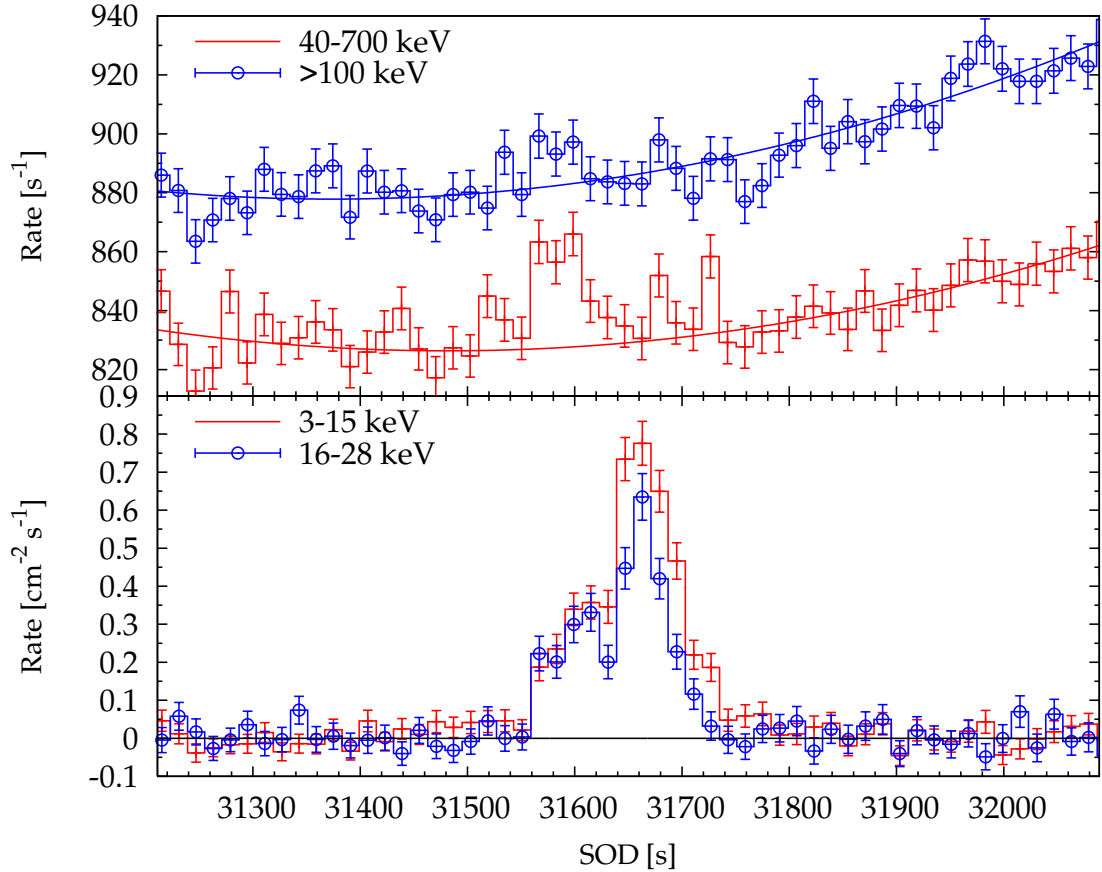}
  \caption{{\bf Top}: BeppoSAX/GRBM light curves of GRB~980429 in two energy channels: 
	40--700\,keV (red crosses) and $>100$\,keV (blue circles). 
	The binning time is 16\,s.
	Time is expressed in seconds of day (SOD).
	The solid lines show the background level, resulted from parabolic
	interpolation.
	{\bf Bottom}: BeppoSAX/WFC light curves in the 3--15 (red crosses) 
	and 16--28\,keV (blue circles) energy bands. 
	\label{grb980429:fig_sax_lcv} }
\end{figure}

Prompted by this finding, we performed a reanalysis of the  BeppoSAX/GRBM
data on this transient.
We discovered that the  X-ray Fast Transient in 
\cite{Heise:1998IAUC6892} is an X-ray rich gamma-ray burst, instead. 
The data were reduced and analyzed following the procedures described in \cite{Guidorzi:phd}. 
Fig.~\ref{grb980429:fig_sax_lcv} 
(top) shows the light curves in the two energy channels  
40--700\,keV and $>100$\,keV. The detection is $7.4$ and $3.3$ $\sigma$ significant
in the lower and higher energy channels, respectively.

\grbsource{} is a $\sim200$\,s long GRB, with an onset time
of 1998-04-29 08:46 UT.   
The GRBM spectrum was fit with a simple power law.  
We find a total 40--700 keV fluence of  $1.7_{-0.4}^{+0.5} \times 10^{-6}$
erg cm$^{-2}$ and a photon index of $3.4_{-0.6}^{+0.7}$.  
The peak flux evaluated over 16\,s (40--700 keV) is 
$(2.7\pm0.9)\times 10^{-8}$ erg cm$^{-2}$ s$^{-1}$ with a photon
index of $2.5\pm0.8$.   
We also re-analysed the data collected by the WFCs. The analysis was 
performed with the BeppoSAX WFC Data Analysis System (version 204).
Fig.~\ref{grb980429:fig_sax_lcv} 
(bottom) shows the 3--15 and 16--28\,keV light curves.

\begin{theacknowledgments}
We thank the Swift team for making these observations possible,
in particular the duty scientists and science planners. 
We acknowledge contract ASI/INAF I/023/05/0. 
PR thanks INAF-IASFMi, where most of the work was carried out, 
for their kind hospitality. 

\end{theacknowledgments}


\end{document}